\documentclass[letterpaper, journal, 10pt]{IEEEtran}

\usepackage{etex}

\usepackage{ifpdf}
\ifpdf
\usepackage[draft,pdfborder={0 0 0}]{hyperref}
\fi

\usepackage{cite}

\usepackage{graphicx}
\usepackage{multirow}

\usepackage[cmex10]{amsmath}
\usepackage{amssymb}
\usepackage{nicefrac}

\usepackage{fixltx2e}

\usepackage{booktabs}
\usepackage{dcolumn}
\usepackage{tabularx}

\usepackage{bm}
\usepackage[varg]{txfonts}
\let\mathbb=\varmathbb
\DeclareSymbolFont{letters}{OML}{ztmcm}{m}{it}

\usepackage{tikz,pgf}
\usepackage{pgfplots}
\usetikzlibrary{shapes,positioning,arrows,decorations.markings,fit,calc,patterns}

\usepackage{algorithm}
\usepackage{algorithmicx}
\usepackage{algpseudocode}
\algrenewcommand{\algorithmiccomment}[1]{\textbf{//}#1}

\usepackage{subfig}

\usepackage{color}

\newcommand{\mvec}[1]{\bm{#1}}

\newcommand{\sgn}[1]{\text{sgn}(#1)}

\newcolumntype{C}[1]{>{\centering\arraybackslash}m{#1}}
\newcolumntype{R}[1]{>{\raggedleft\arraybackslash}m{#1}}

\title{Fast Low-Complexity Decoders\\ for Low-Rate Polar Codes}

\author{Pascal Giard, Alexios Balatsoukas-Stimming, Gabi Sarkis, Claude Thibeault, and Warren J. Gross\\
\thanks{P. Giard, G. Sarkis and W. J. Gross are with the Department of Electrical and Computer Engineering, McGill University, Montr\'eal, Qu\'ebec, Canada (e-mail: \{pascal.giard,gabi.sarkis\}@mail.mcgill.ca, warren.gross@mcgill.ca).}%
\thanks{A. Balatsoukas-Stimming is with the Telecommunications Circuits Laboratory, \'Ecole polytechnique f\'ed\'erale de Lausanne, Lausanne, Switzerland (e-mail: alexios.balatsoukas@epfl.ch).}%
\thanks{C. Thibeault is with the Department of Electrical Engineering, \'Ecole de technologie sup\'erieure, Montr\'eal, Qu\'ebec, Canada (e-mail: claude.thibeault@etsmtl.ca).}}

\newcounter{myEnumCounter}

\begin{document}

\maketitle

\begin{abstract}
Polar codes are capacity-achieving error-correcting codes with an explicit construction that can be decoded with low-complexity algorithms. In this work, we show how the state-of-the-art low-complexity decoding algorithm can be improved to better accommodate low-rate codes. More constituent codes are recognized in the updated algorithm and dedicated hardware is added to efficiently decode these new constituent codes. We also alter the polar code construction to further decrease the latency and increase the throughput with little to no noticeable effect on error-correction performance. Rate-flexible decoders for polar codes of length $1024$ and $2048$ are implemented on FPGA. Over the previous work, they are shown to have from $22$\% to $28$\% lower latency and $26$\% to $34$\% greater throughput when decoding low-rate codes. On 65~nm ASIC CMOS technology, the proposed decoder for a $(1024, 512)$ polar code is shown to compare favorably against the state-of-the-art ASIC decoders. With a clock frequency of 400 MHz and a supply voltage of 0.8 V, it has a latency of 0.41 $\mu$s and an area efficiency of 1.8 Gbps/mm$^2$ for an energy efficiency of 77 pJ/info. bit. At 600 MHz with a supply of 1 V, the latency is reduced to 0.27 $\mu$s and the area efficiency increased to 2.7 Gbps/mm$^2$ at 115 pJ/info. bit.
\end{abstract}

\begin{IEEEkeywords}Polar codes, successive-cancellation decoding\end{IEEEkeywords}

\section{Introduction}
\label{sec:introduction}

Polar codes provably achieve the capacity of binary symmetric memoryless channels when decoded with the low-complexity successive-cancellation (SC) algorithm \cite{Arikan2009}.
The capacity-achieving property and the low-complexity SC decoding have spurred a significant interest in the field leading to the implementation of many hardware decoders based on the SC algorithm.
However, the SC decoding algorithm is sequential in nature and leads to high-latency, low-throughput decoder implementations.

To increase decoding speed, two new decoding algorithms derived from SC were introduced~\cite{Alamdar-Yazdi2011,Sarkis2014}. These two algorithms work by using dedicated, parallel decoding algorithms on parts of a polar code. They exploit the recursive construction of polar codes and the a priori knowledge of the code structure.

While the Fast-SSC \cite{Sarkis2014} algorithm represents a significant improvement over the previous decoding algorithms, the work in \cite{Sarkis2014} and the optimization presented therein targeted high-rate codes. Modifications of the Fast-SSC algorithm and its hardware implementation targeting low-rate codes were presented in~\cite{Giard_SIPS_2015}. In this paper, we formalize and improve the code construction alteration process used in~\cite{Giard_SIPS_2015} and we present more results using the proposed methods, algorithms and implementation. These results show a $22$\% to $28$\% latency reduction and a $22$\% to $28$\% throughput improvement with little to negligible coding loss for low-rate moderate-length polar codes.

The rest of this paper is organized as follows. Section~\ref{sec:background} provides the necessary background on polar coding. Section~\ref{sec:code_construction} then discusses polar code construction alteration along with our proposed method leading to improved latency and throughput of a hardware decoder. In Section~\ref{sec:algo}, modifications to the original Fast-SSC algorithms are proposed in order to further reduce the latency and increase the decoding throughput. Sections~\ref{sec:impl} and \ref{sec:results} present the implementation details along with the detailed results on FPGA. Section~\ref{sec:results} also provides ASIC results for our proposed decoder decoding a $(1024, 512)$ polar code for a comparison against state-of-the-art ASIC decoders from the literature. Finally, Section~\ref{sec:conclusion} concludes this paper.

\section{Background}
\label{sec:background}

\subsection{Polar Codes}
\label{sec:bg:codes}
Polar codes are linear block codes with a recursively constructed generator matrix. In other words, a polar code of length $N$ is built from the concatenation of two constituent polar codes of length $N_v=\nicefrac{N}{2}$. This recursive construction is carried out in a way that polarizes the probability of correctly estimating bits: some bit estimates become more reliable and others becomes less reliable. As the blocklength grows to infinity, some bit estimates become fully reliable and the rest become completely unreliable. In an $(N, k)$ polar code, the $k$ most reliable bit are chosen to store information bits. The remaining bits are called frozen bits and are set to predetermined values. 

The encoding process can be described as follows. A vector containing the $k$ information bits is first expanded into a vector of length $N$ by inserting frozen bits at the appropriate frozen bit locations. The resulting vector is then multiplied by the generator matrix $F_N$ to yield the polar codeword, where $F_N = F_2^{\otimes \log_2 N}$, $F_2 = \left[ \begin{smallmatrix} 1 & 0 \\ 1 & 1\end{smallmatrix} \right]$ and $^{\otimes}$ is the Kronecker power~\cite{Arikan2009}. Frozen bits are usually set to zero and their optimal locations depend on the channel type and condition, as discussed in~\cite{Tal2011a}.

It was shown in \cite{Arikan2011} that polar codes can be encoded and decoded systematically. This leads to an improved bit-error rate (BER) without affecting the frame-error rate (FER). In this work, systematic polar codes are used.

\subsection{Successive-Cancellation Decoding}
\label{sec:bg:sc}
A successive-cancellation decoder estimates the bits $u_0,..., u_i$ sequentially using channel information and previously estimated bits. It utilizes three operations: $F$, $G$, and \textsl{Combine}. The $F$ operation, corresponding to a $\oplus$ in Fig.~\ref{fig:pc8}, accepts two log-likelihood ratio (LLR) inputs and provides an LLR output. $G$, corresponding to a $\bullet$ in Fig.~\ref{fig:pc8}, accepts two LLRs and a bit-estimate to generate one LLR. Finally \textsl{Combine}, which comprises both $\oplus$ and $\bullet$ in a left-to-right direction, takes two bit estimates and combines them to provide two bit estimates. These operations are discussed in detail in Section~\ref{sec:bg:fast-ssc}.

\subsection{From Decoder Graph to Decoder Trees}\label{sec:bg:trees}

\begin{figure}[h]
  \centering
  \subfloat[Graph]{\label{fig:pc8}\newcommand{\ubit}[1]{$u_{#1}$}
\newcommand{\fbit}[1]{\color{gray}$u_{#1}$}
\newcommand{\ucw}[1]{$x_{#1}$}
\newcommand{\fcw}[1]{\color{gray}$x_{#1}$}
\newcommand{\ub}[1]{$#1$}
\newcommand{\fb}[1]{\color{gray}$#1$}

\begin{tikzpicture}

\usetikzlibrary{shapes,positioning,arrows,decorations.markings,fit}

\definecolor{varnode_fill}{RGB}{0,0,0}
\definecolor{chknode_fill}{RGB}{255,255,255}

\tikzset{
  chknode/.style={draw,fill=chknode_fill,circle,minimum size=0.3cm, inner sep=0},
  varnode/.style={draw,fill=varnode_fill,circle,minimum size=0.1cm, inner sep=0},
  channel/.style={draw,fill=white,rectangle},
  sep/.style={rectangle,minimum width=0.25cm, inner sep=0},
  empty/.style={rectangle, inner sep=0},
  bit/.style={circle, inner sep = 0}
}

\tikzset{blue dotted/.style={draw=blue!50!white, line width=1pt,
    dash pattern=on 4pt off 4pt,
    inner sep=0.5mm, rectangle, rounded corners}};

\matrix[row sep=1mm, column sep=1mm] {
  \node[bit] (n0s0) {\fb{u_0}}; & \node[chknode] (n0s1) {$+$}; & \node[sep] (s10) {}; & \node[chknode] (n0s2) {$+$}; & \node[empty] {};              & \node[sep] (s20) {}; & \node[chknode] (n0s3) {$+$}; & \node[empty] {}; & \node[empty] {}; & \node[empty] {}; && \node[bit] (xn0s4) {\ub{x_0}};\\
  \node[bit] (n1s0) {\fb{u_1}}; & \node[varnode] (n1s1) {};    & \node[sep] (s11) {}; &                              & \node[chknode] (n1s2) {$+$};  & \node[sep] (s21) {}; & \node[empty] {};             & \node[chknode] (n1s3) {$+$}; & \node[empty] {}; & \node[empty] {}; && \node[bit] (xn1s4) {\ub{x_1}};\\
  \node[bit] (n2s0) {\fb{u_2}}; & \node[chknode] (n2s1) {$+$}; & \node[sep] (s12) {}; & \node[varnode] (n2s2) {};    & \node[empty] {};              & \node[sep] (s22) {}; & \node[empty] {};             & \node[empty] {}; & \node[chknode] (n2s3) {$+$}; & \node[empty] {}; && \node[bit] (xn2s4) {\ub{x_2}};\\

  \node[bit] (n3s0) {\fb{u_3}}; & \node[varnode] (n3s1) {};    & \node[sep] (s13) {}; & \node[empty] {};             & \node[varnode] (n3s2) {};     & \node[sep] (s23) {}; & \node[empty] {};             & \node[empty] {}; & \node[empty] {}; & \node[chknode] (n3s3) {$+$}; && \node[bit] (xn3s4) {\ub{x_3}};\\

  \node[bit] (n4s0) {\fb{u_4}}; & \node[chknode] (n4s1) {$+$}; & \node[sep] (s14) {}; & \node[chknode] (n4s2) {$+$}; & \node[empty] {};              & \node[sep] (s24) {}; & \node[varnode] (n4s3) {};    & \node[empty] {}; & \node[empty] {}; & \node[empty] {}; && \node[bit] (xn4s4) {\ub{x_4}};\\
  \node[bit] (n5s0) {\ub{u_5}}; & \node[varnode] (n5s1) {};    & \node[sep] (s15) {}; &                              & \node[chknode] (n5s2) {$+$};  & \node[sep] (s25) {}; & \node[empty] {};             & \node[varnode] (n5s3) {}; & \node[empty] {}; &  \node[empty] {}; && \node[bit] (xn5s4) {\ub{x_5}};\\
  \node[bit] (n6s0) {\ub{u_6}}; & \node[chknode] (n6s1) {$+$}; & \node[sep] (s16) {}; & \node[varnode] (n6s2) {};    & \node[empty] {};              & \node[sep] (s26) {}; & \node[empty] {};             & \node[empty] {}; & \node[varnode] (n6s3) {}; &  \node[empty] {}; && \node[bit] (xn6s4) {\ub{x_6}};\\
  
  \node[bit] (n7s0) {\ub{u_7}}; & \node[varnode] (n7s1) {};    & \node[sep] (s17) {}; &                              & \node[varnode] (n7s2) {};  & \node[sep] (s27) {}; & \node[empty] {};             & \node[empty] {}; & \node[empty] {}; &  \node[varnode] (n7s3) {}; && \node[bit] (xn7s4) {\ub{x_7}};\\
};
\path[-] (n0s0) edge (n0s1) (n0s1) edge (n0s2) (n0s2) edge (n0s3) (n0s3) edge (xn0s4);
\path[-] (n1s0) edge (n1s1) (n1s1) edge (n1s2) (n1s2) edge (n1s3) (n1s3) edge (xn1s4);
\path[-] (n2s0) edge (n2s1) (n2s1) edge (n2s2) (n2s2) edge (n2s3) (n2s3) edge (xn2s4);
\path[-] (n3s0) edge (n3s1) (n3s1) edge (n3s2) (n3s2) edge (n3s3) (n3s3) edge (xn3s4);
\path[-] (n4s0) edge (n4s1) (n4s1) edge (n4s2) (n4s2) edge (n4s3) (n4s3) edge (xn4s4);
\path[-] (n5s0) edge (n5s1) (n5s1) edge (n5s2) (n5s2) edge (n5s3) (n5s3) edge (xn5s4);
\path[-] (n6s0) edge (n6s1) (n6s1) edge (n6s2) (n6s2) edge (n6s3) (n6s3) edge (xn6s4);
\path[-] (n7s0) edge (n7s1) (n7s1) edge (n7s2) (n7s2) edge (n7s3) (n7s3) edge (xn7s4);

\path[-] (n0s1) edge (n1s1);
\path[-] (n2s1) edge (n3s1);
\path[-] (n4s1) edge (n5s1);
\path[-] (n6s1) edge (n7s1);

\path[-] (n0s2) edge (n2s2);
\path[-] (n1s2) edge (n3s2);
\path[-] (n4s2) edge (n6s2);
\path[-] (n5s2) edge (n7s2);

\path[-] (n0s3) edge (n4s3);
\path[-] (n1s3) edge (n5s3);
\path[-] (n2s3) edge (n6s3);
\path[-] (n3s3) edge (n7s3);

\node (g_n1s2) [blue dotted, fit = (n4s2) (n5s2) (n6s2) (n7s2)] {};

\end{tikzpicture}}
  \subfloat[Decoder Tree]{\label{fig:ssc-tree}\rotatebox{90}{\begin{tikzpicture}[baseline=(base),
        level/.style={level distance = 6mm},
        level 1/.style={sibling distance=19mm, edge from parent/.style={draw,black,line width=2pt}},
        level 2/.style={sibling distance=9mm, edge from parent/.style={draw,black,line width=1pt}},
        level 3/.style={sibling distance=4mm, edge from parent/.style={draw,black,line width=0.5pt}},
        ]

\tikzset{
frozen/.style={thick,draw=black,fill=white,minimum size=3mm,circle, inner sep=0},
fullspace/.style={thick,draw=black,fill=black,minimum size=3mm,circle, inner sep = 0},
mixed/.style={thick,draw=black,fill=gray,minimum size=3mm,circle, inner sep = 0},
ml_mixed/.style={thick,draw=black,fill=blue,minimum size=3mm,circle, inner sep = 0}
}

\tikzset{blue dotted/.style={draw=blue!50!white, line width=1pt,
    dash pattern=on 4pt off 4pt,
    inner sep=0.5mm, rectangle, rounded corners}};

\node[mixed] (3_0){} [grow=left]
	child {node[frozen, label={[font=\scriptsize]right:\rotatebox{-90}{\bf left}}] (2_0){}
	}
	child {node[mixed, label={[font=\scriptsize]right:\rotatebox{-90}{\bf right}}] (2_1){}
		child {node[mixed] (1_2){}
			child {node[frozen] (0_4){}
			}
			child {node[fullspace] (0_5){}
			}
		}
		child {node[fullspace] (1_3){}
		}
	}
;

\node at ($(2_0)+(-0.375,0)$) {\rotatebox{-90}{\fontsize{8}{8}\selectfont $u_0^3$}};
\node at ($(0_4)+(-0.325,0)$) {\rotatebox{-90}{\fontsize{8}{8}\selectfont $u_4$}};
\node at ($(0_5)+(-0.325,0)$) {\rotatebox{-90}{\fontsize{8}{8}\selectfont $u_5$}};
\node at ($(1_3)+(-0.375,0)$) {\rotatebox{-90}{\fontsize{8}{8}\selectfont $u_6^7$}};

\node (g_concat) [blue dotted, fit = (2_1)] {};

\node [circle, below= 5mm of 1_3.base] (base) {};

\end{tikzpicture}}}
  \caption{Diffrent representations of polar codes.}\label{fig:pc8all}
\end{figure}
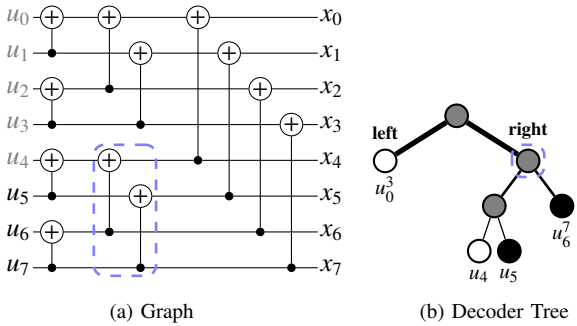

As shown in \cite{Alamdar-Yazdi2011,Sarkis2014}, polar codes can be represented as trees. Using only the three node types introduced in \cite{Alamdar-Yazdi2011}---rate-0, rate-1 and rate-$R$---the graph of Fig.~\ref{fig:pc8} can be represented as the decoder tree of Fig.~\ref{fig:ssc-tree}. The black and gray $u_i$ labels of Fig.~\ref{fig:pc8} correspond to information and frozen bits, respectively.

In Fig.~\ref{fig:ssc-tree}, the rate-0 nodes are white and correspond to a constituent code that contains only frozen bits, while rate-1 nodes are black and correspond to a constituent code that contains only information bits. Finally, the gray nodes are of rate $R$, where $0 < R < 1$, and correspond to a constituent code that contains both frozen and information bits.

\subsection{Fast-SSC Decoding}\label{sec:bg:fast-ssc}

The Fast-SSC algorithm presented in \cite{Sarkis2014} further trims the decoder tree by using different, more efficient algorithms to decode some constituent codes of rate $R$. The constituent codes directly decoded in the Fast-SSC algorithm have different maximum lengths $N_v$ depending on the complexity of the corresponding decoding algorithm. For example, for a repetition node, $N_v$ was set to $16$ while it was set to $4$ for an exhaustive search maximum-likelihood (ML) node.

The decoder tree node types corresponding to the decoding algorithm of \cite{Sarkis2014} are summarized in Table~\ref{tab:node-types}. Note that the ``01'' node was called ``ML'' in \cite{Sarkis2014}. We briefly review the most important operations and leaf node types below.

\subsubsection{F Operations} The $F$ operation generates the messages to be sent to a left child and is performed using the min-sum approximation as defined in~\cite{Leroux2011}:
\begin{equation}\label{eqn:sc:f}
\begin{split}
\alpha_l[i] & = F( \alpha_v[i], \alpha_v[i + \nicefrac{N_v}{2}] )\\
& = \sgn{\alpha_v[i]}\sgn{\alpha_v[i + \nicefrac{N_v}{2}]} \min(|\alpha_v[i]|, |\alpha_v[i + \nicefrac{N_v}{2}]|),
\end{split}
\end{equation}
where $\alpha_v$ are soft reliability values from the parent node, represented as LLRs, and $N_v$ is the node input length.

\subsubsection{G and G\_0R Operations} The $G$ operation generates the messages to be sent to a right child node. It is performed as defined in~\cite{Leroux2011}:
\begin{equation}\label{eqn:sc:g}
\begin{split}
\alpha_r[i] & = G( \alpha_v[i], \alpha_v[i + \nicefrac{N_v}{2}], \beta_l[i] )\\
& = \begin{cases}
\alpha_v[i + \nicefrac{N_v}{2}] + \alpha_v[i]\text{,} & \text{when } \beta_l[i] = 0;\\
\alpha_v[i + \nicefrac{N_v}{2}] - \alpha_v[i]\text{,} & \text{otherwise},
\end{cases}
\end{split}
\end{equation}
where $\beta_l$ is the bit estimate vector generated by the left sibling in the subtree.

The $G\_0R$ operation is a special case of the $G$ operation, where the left-hand-side sibling in the subtree is a rate-0 node, i.e. $\beta_l$ is a all-zero vector.

\subsubsection{Combine and Combine\_0R Operations} The $Combine$ operation corresponds to the concatenation of two bit-estimate vectors generated from the left and right child nodes. As an example, going up the tree, the node circled in blue in Fig.~\ref{fig:ssc-tree} combines the bit-estimate vectors from its children to provide the root node of the decoder tree with a bit estimate vector for the right-hand-side subtree. In the graph representation illustrated in Fig.~\ref{fig:pc8}, the same operation is also circled in blue and illustrates how the $Combine$ operation calculates the bit estimate vector $\beta_v$:
\begin{equation}\label{eqn:sc:combine}
\beta_v[i] = \begin{cases}
\beta_l[i] \oplus \beta_r[i]\text{,} & \text{when } i < \nicefrac{N_v}{2};\\
\beta_r[i - \nicefrac{N_v}{2}]\text{,} & \text{otherwise,}
\end{cases}
\end{equation}
where $\beta_l$ and $\beta_r$ are bit estimates emanating from the left and right child nodes, respectively.

Similar to the $G\_0R$ operation, the $Combine\_0R$ operation is a special case where the left-hand-side sibling in the subtree is a rate-0 node i.e. $\beta_l$ is a all-zero vector.

\subsubsection{Repetition Nodes} Repetition nodes provide the bit estimate vector for a repetition code. A repetition code contains a single information bit that is replicated over the length $N_v$ of the code. The maximum-likelihood decoding rule for these codes is to perform threshold detection on the sum of the input LLRs~\cite{Sarkis2014}.

\subsubsection{Single-parity-check Nodes} Single-parity-check (SPC) nodes provide bit estimates for SPC codes. The bits in an SPC codeword satisfy the even parity constraint, i.e. they sum to $0$ under binary addition. The maximum likelihood algorithm to decode them consists of first verifying that hard decisions on the input reliability values satisfy the parity contraint. In case where they do not, the hard decision of the least reliable input bit is flipped.

\subsubsection{Other Nodes}
The remaining nodes are a concatenation of the operations and nodes described above, or an extension. For example, the RepSPC node is the concatenation of a repetition node with an SPC node with a constant length $N_v=8$. Another example is the 0SPC node, which is the concatenation of a rate-0 node and an SPC node. Thus, it replaces the execution of three operations: a $G\_0R$ operation, an SPC operation and a $Combine\_0R$ operation.

\begin{table}[t]
\centering
\setlength{\tabcolsep}{3pt}
\caption{Decoder tree node types supported by the original Fast-SSC polar decoder.}
\begin{tabularx}{\columnwidth}{ccp{4.5cm}}
  \toprule
  \textbf{Name} & \textbf{Color} & \textbf{Description}\\
  \midrule
  0R     & White and gray   & Left-half side is frozen.\\
  R1     & Gray and black   & Right-half side is all information.\\
  RSPC   & Gray and yellow  & Right-half side is an SPC code.\\
  0SPC   & White and yellow & Left-half side is frozen, right-half side is an SPC code.\\
  Rep    & Green            & Repetition code, maximum length $N_v$ of 16.\\
  RepSPC & Green and yellow & Concatenation of a repetition code on the left and an SPC code on the right, $N_v = 8$.\\
  01     & Black and white  & Fixed-length pattern $N_v=4$ where the left-half side is frozen and the right-half side is all information.\\
 rate-$R$& Gray             & Mixed rate node.\\
  \bottomrule
\end{tabularx}
\label{tab:node-types}
\end{table}

With the specialized nodes and operations of Fast-SSC, the trimmed decoder tree for the $(8,3)$ polar code illustrated in Fig.~\ref{fig:pc8all} is made of a single 0SPC node.

\section{Altering the Code Construction}\label{sec:code_construction}


\subsection{Original Construction}\label{sec:code_construction:orig}
\begin{figure}[t]
  \centering
  \includegraphics[width=\columnwidth]{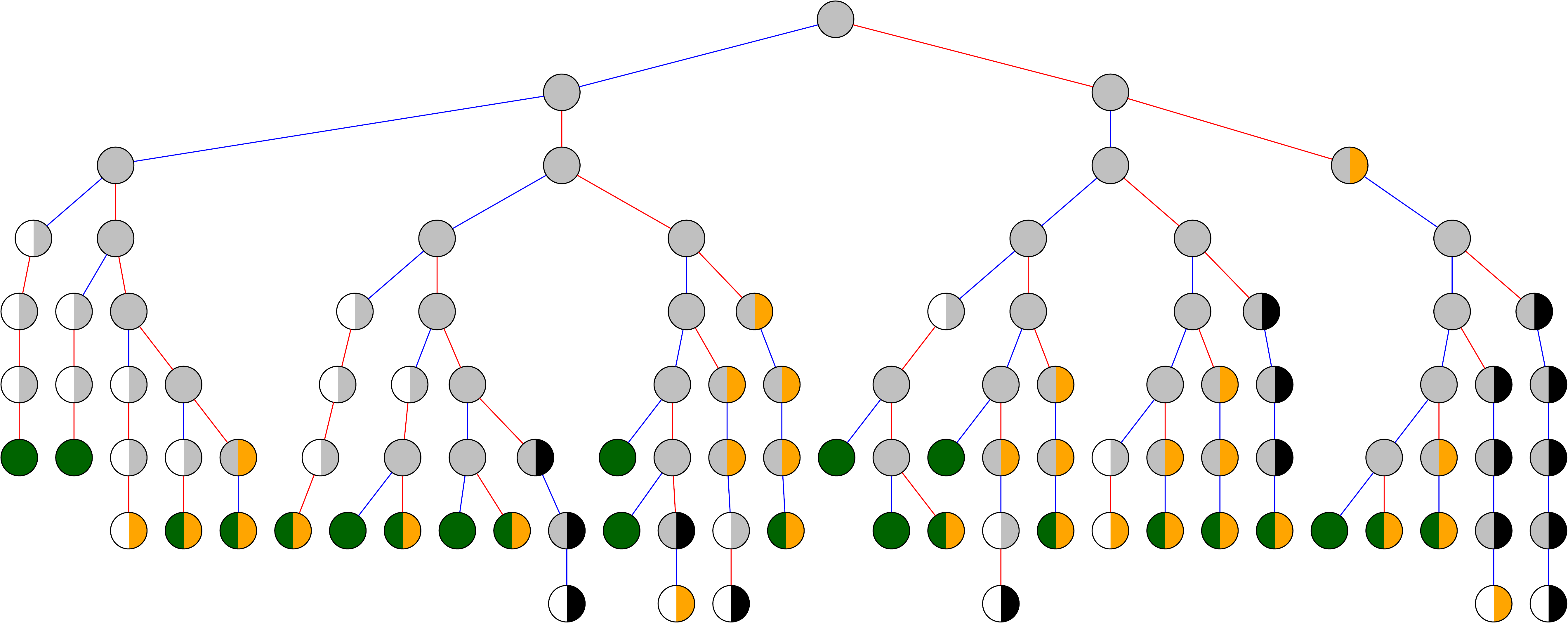}
  \caption{Decoder tree for the (1024, 512) polar code built using \cite{Tal2011a} and decoded with the nodes and operations of  \cite{Sarkis2014}.}
  \label{fig:tree-fast-ssc-s075}
\end{figure}

As mentioned in Section~\ref{sec:bg:codes}, a good polar code is constructed by selecting which bits to freeze, according to the type of channel and its conditions \cite{Arikan2009,Mori2009,Tal2011a,Trifonov2012}. Fig.~\ref{fig:tree-fast-ssc-s075} shows the decoder tree corresponding to the $(1024,512)$ polar code constructed using the technique of \cite{Tal2011a} where only the node types defined in Table~\ref{tab:node-types} are used with the same constraints of \cite{Sarkis2014}. The polar code was optimized for an $\nicefrac{E_b}{N_0}$ of 2.5~dB. The $F$, $G$, $G\_0R$, $Combine$ and $Combine\_0R$ blocks are constrained to a maximum of $P=512$ inputs meaning that, for nodes with a length $N_v > P$, $\left \lceil \nicefrac{N_v}{P} \right \rceil$ cycles are required. The $Rep$, $RepSPC$ and $01$ blocks are all executed in one clock cycle. Finally, the SPC-based nodes---0SPC and RSPC---use pipelining and require $\left \lceil \nicefrac{N_v}{P} \right \rceil+4$ clock cycles. Thus the decoding latency to decode the tree of Fig.~\ref{fig:tree-fast-ssc-s075} using the algorithm and implementation of \cite{Sarkis2014} is 220 clock cycles (CC) and the information throughput is 2.33~bits/CC.

Altering a polar code to further trim the decoder tree can result in a significant latency reduction, without affecting the code rate. By making these modifications however, the error-correction performance is degraded.
Although, as will be shown in the next section, the impact can be small, especially if the number of changes is limited.

\subsection{Altered Polar Code Construction}
In all simplified SC decoders, the size of the decoder tree depends on the distribution of frozen and information bit locations in the code. Ar{\i}kan's original polar code construction only focuses on maximizing the reliability of the information bits. Several altered polar-like code constructions have been proposed in the literature~\cite{Huang2012,Balatsoukas2014,Zhang2015} and their objective is to trade off error-correction performance for decoding complexity by slightly changing the set of information bits, while keeping the code rate fixed. The main idea behind all the altered code constructions is to exchange the locations of a few frozen bits and information bits in order to get more bit patterns that are favorable in terms of decoding latency. In all cases, care must be taken in order to avoid using bit locations that are highly unreliable to transmit information bits.

The method in \cite{Huang2012} first defines a small set of bit locations which contains the $n_s-h$ least reliable information bit locations along with the $h$ most reliable frozen bit locations. Then, in order to keep the rate fixed, it performs an exhaustive search over all $n_s \choose h$ possible combinations of the $n_s$ elements containing exactly $h$ frozen bit locations and selects the combination that leads to the smallest decoding latency. In \cite{Balatsoukas2014}, the altered construction problem is formalized as a binary integer linear program. Consequently, it is shown that finding the polar code with the lowest decoding complexity under an error-correction performance constraint is an NP-hard problem. For this reason, a greedy approximation algorithm is presented which provides reasonable results at low complexity even for large code lengths. A similar greedy algorithm is presented in \cite{Zhang2015} for polar codes with more general code lengths of the form $N = l^n,$ $l \geq 2$.

\subsection{Proposed Altered Construction}
The methods of \cite{Balatsoukas2014,Zhang2015} only considered rate-$0$ and rate-$1$ nodes. As such, the results can not be directly applied to Fast-SSC decoding, where several additional types of special nodes exist. For this reason, in this work we follow the more general exhaustive search method of \cite{Huang2012}, augmented with a human-guided approach. 

More specifically, bit-state alterations that would lead to smaller latency are identified by visual inspection of the decoder tree for the unaltered polar code. This list of bit locations is then passed to a program to be added to the bit locations considered by the technique described in \cite{Huang2012}. Hence, two lists are composed: one that contains frozen bit locations proposed by the user as well as locations that were almost reliable enough to be used to carry information bits, and one that contains the information bit locations proposed by the user and the locations that barely made it into information bit locations.

The code alteration algorithm then proceeds by gradually calculating the decoding latency for all possible bit swap combinations. A constrained-size and ordered list of the combinations with the lowest decoding latency is kept. Once that list needs to be trimmed, only one entry per latency value is kept by simulating the error-correction performance of the altered code at an $\nicefrac{E_b}{N_0}$ value of interest. The entry with the best frame-error rate is kept and the others with the same latency are removed from the list. That list containing the best candidates is further trimmed by removing all candidates that feature both a greater latency and worse error-correction performance compared to those of their predecessor. Similarly to the technique of \cite{Huang2012}, our proposed technique does not alter the code rate as the total number of information and frozen bits remains the same.

\subsubsection{Human-guided Criteria}
The suggested bits to swap are selected to improve the latency and throughput. Thus, these bit swaps must eliminate constituent codes for which we do not have an efficient decoding algorithm and create ones for which we do. We classify the selection criteria under two categories: the bit swaps that transform frozen bit locations into information bit locations and bit swaps that do the opposite. The former increase the coding rate while the latter reduce it.

In addition to the node type definitions of Table~\ref{tab:node-types}, the below descriptions of criteria use the following types of subtrees or nodes:
\begin{itemize}
  \item R1-01: subtree rooted in a R1 node with a 01 leaf node, may contain a chain of R1 nodes 
  \item Rep1: subtree rooted in a R1 node with a leaf Rep node; in Section~\ref{sec:algo}, that subtree is made into a node where the left-half side is a repetition code and the right-half side is all information
  \item R1-RepSPC: subtree rooted in a R1 node with a RepSPC leaf node, may contain a chain of R1 nodes
  \item Rep-Rep1: subtree where the rate-$R$ node has a left-hand-side and right-hand-side nodes are Rep and Rep1 nodes, respectively
  \item 0-RepSPC: subtree rooted in a 0R node with a leaf RepSPC node; in Section~\ref{sec:algo}, that subtree is made into a node where the left-half side is frozen and the the right-half side is a RepSPC node
\end{itemize}

Dedicated hardware to efficiently decode Rep1 and 0RepSPC nodes are presented in Section~\ref{sec:algo}.

\subsubsection*{From frozen to information locations}
\begin{enumerate}
  \item Unfreezing the second bit of a 01 node that is part of a R1-01 subtree creates an RSPC node.
  \item Changing an RepSPC into an RSPC node by adding the second, third and fifth bit locations.
  \item Changing a RSPC node into a R1 node by changing the SPC code into a rate-1 code.
    \setcounter{myEnumCounter}{\value{enumi}}
\end{enumerate}

Criterion $1$ is especially beneficial where the R1-01 subtree contains a chain of R1 nodes, e.g., Pattern 5 in Fig.~\ref{fig:tree-alt-example}. Similarly, Criterion $2$ has a significant impact on R1-RepSPC subtrees containing a chain of R1 nodes, e.g., Pattern $3$ in Fig.~\ref{fig:tree-alt-example}.

\subsubsection*{From information to frozen locations}
\begin{enumerate}
  \setcounter{enumi}{\value{myEnumCounter}}
  \item Changing a 0R-01 subtree into a Repetition node.
  \item Freezing the only information bit location of a Rep node to change it into a rate-0 code.
  \item A specialization of the above, changing a Rep-RepSPC subtree into a 0-RepSPC subtree by changing the left-hand-side Rep node into a rate-0 node.
  \item Transforming a Rep-Rep1 subtree into a 0-RepSPC subtree by changing the left-hand-side repetition code into a rate-0 code and by freezing the fifth bit location of the Rep1 subtree to change the rate-1 code into an SPC code.
\end{enumerate}

\begin{figure}[t]
  \centering
  \def\svgwidth{0.7\columnwidth}
  \input{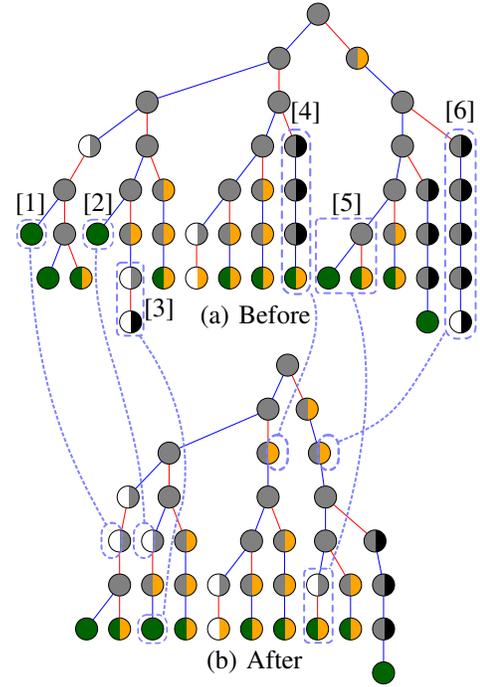}
  \caption{Decoder trees for two different $(512, 376)$ polar codes, where (a) and (b) are before and after construction alteration, respectively.}
  \label{fig:tree-alt-example}
\end{figure}

Consider the decoder tree for a $(512, 376)$ polar code as illustrated in Fig.~\ref{fig:tree-alt-example}a, where some frozen bit patterns are circled in blue and numbered for reference. Its implementation results in a decoding latency of 106 clock cycles. That latency can be significantly reduced by freezing information bit locations or by transforming previously frozen locations into information bits.

Notably, five of the bit-swapping criteria---leading to latency reduction---described above are illustrated in Fig.~\ref{fig:tree-alt-example}a. The patterns numbered $1$ and $2$ are repetition nodes meeting the fourth criterion. Changing both into rate-0 nodes introduces two new 0R nodes. The patterns $3$ to $6$ are illustrations of the fourth, second, sixth and first criteria, respectively. 

Fig.~\ref{fig:tree-alt-example}b shows the resulting decoder tree after the alterations were made. The latency has been reduced from 106 to 82 clock cycles.

\subsubsection{Example Results}
\begin{figure}[t]
  \centering
  \includegraphics[width=\columnwidth]{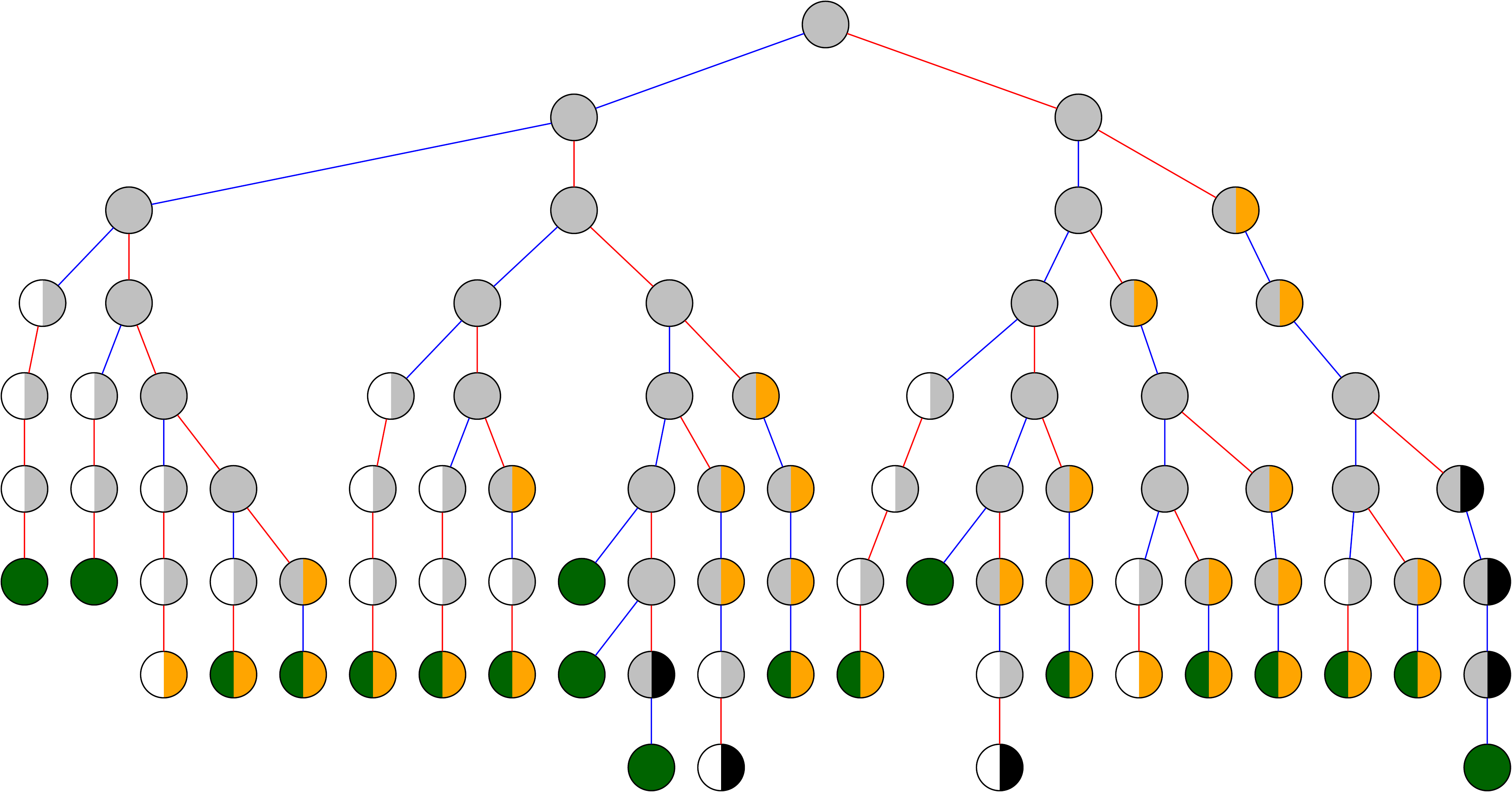}
  \caption{Decoder tree for the altered $(1024, 512)$ polar code.}
  \label{fig:tree-fast-ssc-alt}
\end{figure}

Applying our proposed altered construction method, we were able to decrease the decoding latency of the $(1024, 512)$ polar code illustrated in Fig.~\ref{fig:tree-fast-ssc-s075} from 220 to 189 clock cycles, a 14\% improvement, with 5 bit swaps. That increases the information throughput to 2.71 bits/CC, up from 2.33 bits/CC. The corresponding decoder tree is shown in Fig.~\ref{fig:tree-fast-ssc-alt}.

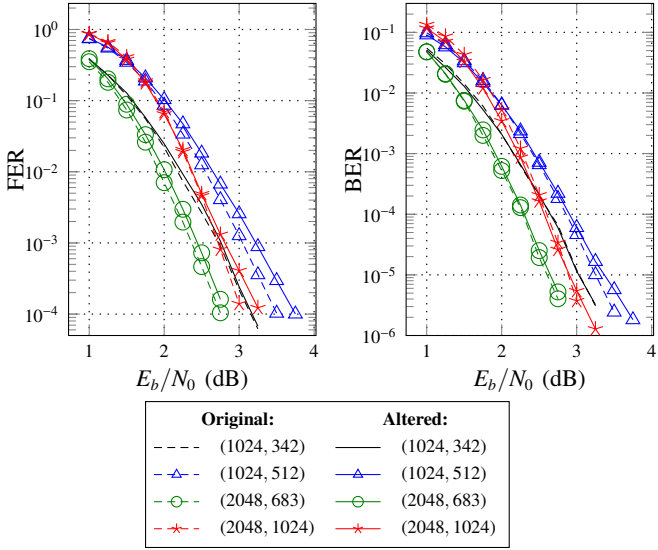
\begin{figure}[t]
  \centering
  \definecolor{darkgreen}{RGB}{0,128,0}

\begin{tikzpicture}

  \pgfplotsset{
    grid style = {
      dash pattern = on 0.05mm off 1mm,
      line cap = round,
      black,
      line width = 0.5pt
    },
    label style = {font=\fontsize{9pt}{7.2}\selectfont},
    tick label style = {font=\fontsize{7pt}{7.2}\selectfont}
  }

  \begin{semilogyaxis}[%
    xlabel=$E_b/N_0$ (dB),%
    xlabel style={yshift=0.6em},%
    ylabel=FER, ylabel style={yshift=-1.5em},%
    ymin=5e-5,
    width=0.55\columnwidth, height=6cm, grid=major,%
    legend style={
      anchor={center},
      cells={anchor=west},
      column sep= 2mm,
      font=\fontsize{7pt}{7.2}\selectfont,
      mark options=solid
    },
    legend to name=perf-alt-legend,
    legend columns=2,
    mark size=3.0pt, mark options=solid]

    \addlegendimage{empty legend}
    \addlegendentry{\hspace{-7pt}\textbf{Original:}}
    \addlegendimage{empty legend}
    \addlegendentry{\hspace{-7pt}\textbf{Altered:}}

    \addplot[color=black,densely dashed] table[x=ebn0_db,y=FER] {n1024_k342_awgn_s0.90.txt};
    \addlegendentry{$(1024, 342)$}
    \addplot[color=black] table[x=ebn0_db,y=FER] {n1024_k342_awgn_s0.90.swap.679.711.791.247.575.to.817.459.621.667.737.txt};
    \addlegendentry{$(1024, 342)$}

    \addplot[color=blue,densely dashed,mark=triangle] table[x=snr_db,y=FER] {1k_s0.75.txt};
    \addlegendentry{$(1024, 512)$}
    \addplot[color=blue,mark=triangle] table[x=snr_db,y=FER] {1k_final_syst.float.txt};
    \addlegendentry{$(1024, 512)$}

    \addplot[color=darkgreen,mark=o,densely dashed] table[x=ebn0_db,y=FER] {n2048_k683_awgn_s0.93.txt};
    \addlegendentry{$(2048, 683)$}
    \addplot[color=darkgreen,mark=o] table[x=ebn0_db,y=FER] {n2048_k683_awgn_s0.93.swap.1799.1671.1367.1567.1583.1255.1359.1151.1335.919.to.1369.1825.937.938.1265.1266.1729.1730.1809.1810.txt};
    \addlegendentry{$(2048, 683)$}

    \addplot[color=red,densely dashed,mark=star] table[x=ebn0_db,y=FER] {n2048_k1024_awgn_s0.708.txt};
    \addlegendentry{$(2048, 1024)$}
    \addplot[color=red,mark=star] table[x=ebn0_db,y=FER] {n2048_k1024_awgn_s0.708.swap.807.431.439.839.903.463.1223.607.671.1295.to.473.1249.1329.1800.1808.1793.1794.1796.1696.1728.txt};
    \addlegendentry{$(2048, 1024)$}

  \end{semilogyaxis}
\end{tikzpicture}
\begin{tikzpicture}

  \pgfplotsset{
    grid style = {
      dash pattern = on 0.05mm off 1mm,
      line cap = round,
      black,
      line width = 0.5pt
    },
    label style = {font=\fontsize{9pt}{7.2}\selectfont},
    tick label style = {font=\fontsize{7pt}{7.2}\selectfont}
  }

  \begin{semilogyaxis}[%
    xlabel=$E_b/N_0$ (dB),%
    xlabel style={yshift=0.6em},%
    ylabel=BER, ylabel style={yshift=-1.5em},%
    ymin=1e-6, ymax=3e-1,%
    width=0.55\columnwidth, height=6cm, grid=major,%
    mark size=3.0pt,mark options=solid]

    \addplot[color=black,densely dashed] table[x=ebn0_db,y=BER] {n1024_k342_awgn_s0.90.txt};
    \addplot[color=black] table[x=ebn0_db,y=BER] {n1024_k342_awgn_s0.90.swap.679.711.791.247.575.to.817.459.621.667.737.txt};

    \addplot[color=blue,densely dashed,mark=triangle] table[x=snr_db,y=BER] {1k_s0.75.txt};
    \addplot[color=blue,mark=triangle] table[x=snr_db,y=BER] {1k_final_syst.float.txt};

    \addplot[color=darkgreen,mark=o,densely dashed] table[x=ebn0_db,y=BER] {n2048_k683_awgn_s0.93.txt};
    \addplot[color=darkgreen,mark=o] table[x=ebn0_db,y=BER] {n2048_k683_awgn_s0.93.swap.1799.1671.1367.1567.1583.1255.1359.1151.1335.919.to.1369.1825.937.938.1265.1266.1729.1730.1809.1810.txt};

    \addplot[color=red,densely dashed,mark=star] table[x=ebn0_db,y=BER] {n2048_k1024_awgn_s0.708.txt};
    \addplot[color=red,mark=star] table[x=ebn0_db,y=BER] {n2048_k1024_awgn_s0.708.swap.807.431.439.839.903.463.1223.607.671.1295.to.473.1249.1329.1800.1808.1793.1794.1796.1696.1728.txt};

  \end{semilogyaxis}
\end{tikzpicture}
\\
\ref*{perf-alt-legend}
  \caption{Error-correction performance using binary phase shift keying over an additive white Gaussian channel of the altered codes compared to that of the original codes constructed using the Tal and Vardy method~\cite{Tal2011a}.}
\label{fig:tal-vs-alt}
\end{figure}

The error-correction performance of the $(1024, 512)$ altered code is degraded as illustrated by the markerless black curves in Fig.~\ref{fig:tal-vs-alt}. The loss amounts to less than 0.25 dB at a FER of $10^{-4}$. For wireless applications, which are usually the target for codes of such lengths and rates, this represents the FER range of interest.

Fig.~\ref{fig:tal-vs-alt} also shows the error-correction performance of three other polar codes altered using our proposed method. In the case of these other codes, the alterations have a negligible effect on error-correction performance

\section{New Constituent Decoders}\label{sec:algo}

Looking at the decoder tree of Fig.~\ref{fig:tree-fast-ssc-alt}, it can be seen that some frozen bit patterns occur often. Adding support for more constituent codes to the Fast-SSC algorithm will result in reduced latency and increased throughput under the constraint that the corresponding computation nodes do not significantly lengthen the critical path of a hardware implementation. As a result of an investigation, the constituent codes of Table~\ref{tab:new-node-types} were added. Furthermore, post-place and route timing analysis showed that the maximum length $N_v$ of a repetition node could be increased from 16 to 32 without affecting the critical path.

\begin{table}[t]
\centering
\setlength{\tabcolsep}{3pt}
\caption{New functions performed by the proposed decoder.}
\begin{tabularx}{\columnwidth}{ccp{4.25cm}}
  \toprule
  \textbf{Name} & \textbf{Color} & \textbf{Description}\\
  \midrule
  Rep1    & Green and black     & Repetition code on the left, rate-1 code on the right, maximum length $N_v$ of 8.\\
  0RepSPC & White and lilac     & Rate-0 code on the left, RepSPC code on the right, $N_v = 16$.\\
  001     & $\frac{3}{4}$ white and $\frac{1}{4}$ black & Rate-0 code on the left, 01 code on the right, $N_v = 8$.\\
  \bottomrule
\end{tabularx}
\label{tab:new-node-types}
\end{table}

\begin{figure}[t]
  \centering
  \includegraphics[width=\columnwidth]{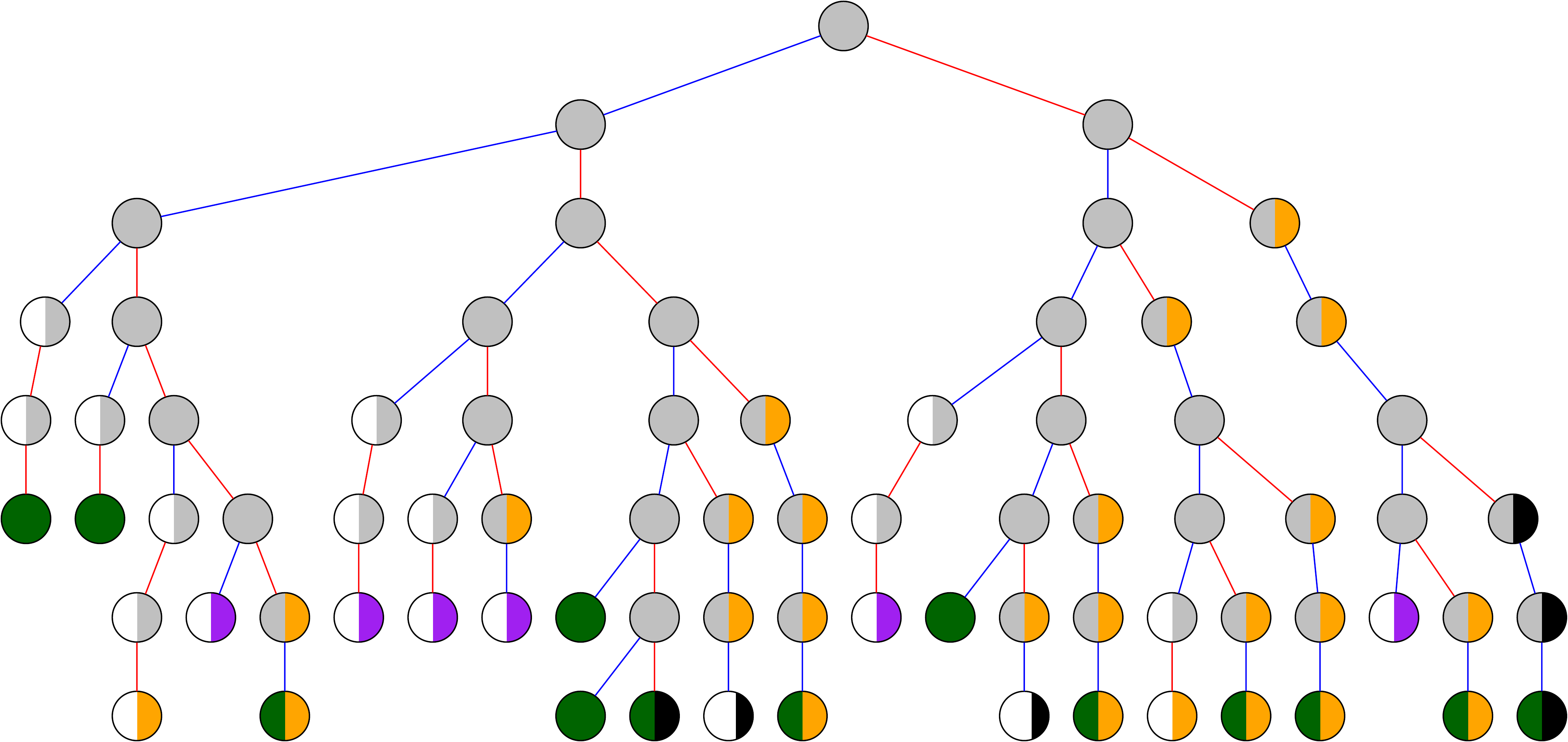}
  \caption{Decoder tree for the altered polar code with the added nodes.}
  \label{fig:tree-fast-ssc-final}
\end{figure}

The new decoder tree shown in Fig.~\ref{fig:tree-fast-ssc-final} has a decoding latency of 165 clock cycles, which is a 13\% improvement over the decoder tree of Fig.~\ref{fig:tree-fast-ssc-alt} decoded with the original Fast-SSC algorithm. Thus, the information throughput of that polar code has been improved to 3.103 bits/CC.

To summarize, Table~\ref{tab:frozen-bit-patterns} lists the frozen bit patterns that can be decoded by leaf nodes. It can be seen that the smallest possible leaf node has length $N_v=4$ while our proposed decoder tree shown in Fig.~\ref{fig:tree-fast-ssc-final} has a minimum length $N_v=8$. In other words, Fig.~\ref{fig:tree-fast-ssc-final} is representative of the patterns listed in Table~\ref{tab:frozen-bit-patterns} but not comprehensive.

\begin{table}[t]
\centering
\caption{Frozen bit patterns decoded by leaf nodes.}
\begin{tabular}{cl}
  \toprule
  \textbf{Name} & \textbf{Pattern}\\
  \midrule
  Rep & 0001\\
      & 0000 0001\\
      & 0000 0000 0000 0001\\
      & 0000 0000 0000 0000 0000 0000 0000 0001\vspace{2pt}\\
  Rep1 & 0001 1111\vspace{2pt}\\
  0SPC & 0000 0111\vspace{2pt}\\
  RepSPC & 0001 0111\vspace{2pt}\\
  0RepSPC & 0000 0000 0001 0111\vspace{2pt}\\
  01 & 0011\vspace{2pt}\\
  001 & 0000 0011\vspace{2pt}\\
  \bottomrule
\end{tabular}
\label{tab:frozen-bit-patterns}
\end{table}

\section{Implementation}\label{sec:impl}

\subsection{Quantization}
\label{sec:impl:qtz}
\begin{figure}[t]
  \centering
  \begin{tikzpicture}

  \pgfplotsset{
    grid style = {
      dash pattern = on 0.05mm off 1mm,
      line cap = round,
      black,
      line width = 0.5pt
    },
    label style = {font=\fontsize{9pt}{7.2}\selectfont},
    tick label style = {font=\fontsize{7pt}{7.2}\selectfont}
  }

  \begin{semilogyaxis}[%
    xlabel=$E_b/N_0$ (dB),%
    xlabel style={yshift=0.6em},%
    ylabel=FER, ylabel style={yshift=-1.5em},%
    width=0.55\columnwidth, height=6cm, grid=major,%
    legend style={
      anchor={center},
      cells={anchor=west},
      column sep= 2mm,
      font=\fontsize{7pt}{7.2}\selectfont,
    },
    legend to name=perf-2-legend,
    legend columns=3,
    mark size=3.0pt]

    \addplot[color=black,mark=pentagon] table[x=snr_db,y=FER] {1k_final_syst.float.txt};
    \addlegendentry{Floating-point}

    \addplot[color=red,mark=triangle] table[x=snr_db,y=FER] {1k_final_syst.651.txt};
    \addlegendentry{Fixed-point 6.5.1}

  \end{semilogyaxis}
\end{tikzpicture}
\begin{tikzpicture}

  \pgfplotsset{
    grid style = {
      dash pattern = on 0.05mm off 1mm,
      line cap = round,
      black,
      line width = 0.5pt
    },
    label style = {font=\fontsize{9pt}{7.2}\selectfont},
    tick label style = {font=\fontsize{7pt}{7.2}\selectfont}
  }

  \begin{semilogyaxis}[%
    xlabel=$E_b/N_0$ (dB),%
    xlabel style={yshift=0.6em},%
    ylabel=BER, ylabel style={yshift=-1.5em},%
    width=0.55\columnwidth, height=6cm, grid=major,%
    mark size=3.0pt]

    \addplot[color=black,mark=pentagon] table[x=snr_db,y=BER] {1k_final_syst.float.txt};

    \addplot[color=red,mark=triangle] table[x=snr_db,y=BER] {1k_final_syst.651.txt};

  \end{semilogyaxis}
\end{tikzpicture}
\\
\ref*{perf-2-legend}
  \caption{Impact of quantization on the error-correction performance of the proposed $(1024, 512)$ polar code.}
\label{fig:perf_quant}
\end{figure}
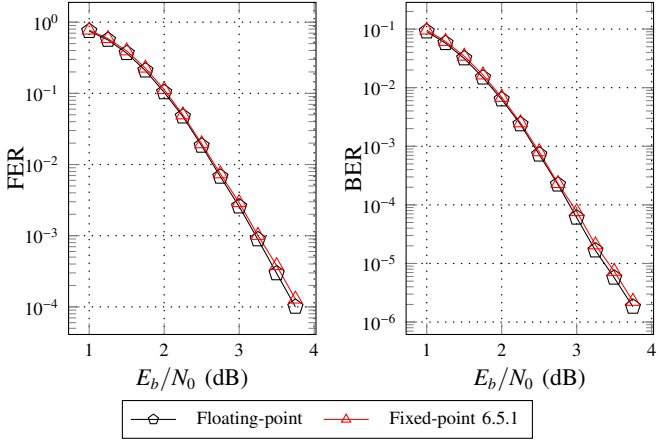

Let $Q_i$ be the total number of bits used to represent LLRs internally, $Q_c$ be the total number of bits to represent channel LLRs, and $Q_f$ be the number of bits among $Q_i$ or $Q_c$ used to represent the fractional part of any LLR. It was found through simulations that using $Q_i.Q_c.Q_f = $ 6.5.1 quantization led to an error-correction performance very close to that of the floating-point number representation for the codes of interest as can be seen in Fig.~\ref{fig:perf_quant}.

\subsection{Rep1 Node}
The Rep1 node decodes Rep1 codes---the concatenation of a repetition code and a rate-1 code---of length $N_v=8$. Its bit-estimate vector $\beta_0^7$ is calculated using operations described in the previous sections. However, instead of performing the required operations sequentially, the dedicated hardware preemptively calculates intermediate soft values.

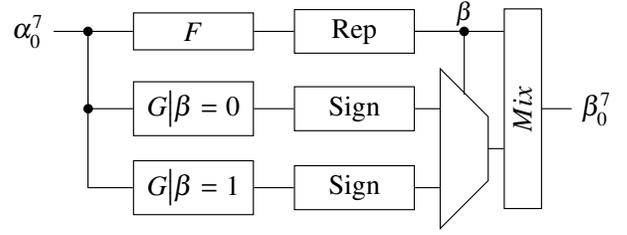
\begin{figure}[t]
  \centering 
  \resizebox{\columnwidth}{!}{
    \begin{tikzpicture}

\usetikzlibrary{shapes,positioning,arrows,decorations.markings,fit,mux}

\tikzset{
  block/.style={draw,rectangle, minimum width = 1.5cm},
  >={latex},
  connector/.style={circle,draw,fill=black,minimum size=3pt,inner sep=0},
}

\matrix[row sep=3mm, column sep=5mm] {

\node[rectangle] (alpha) {$\alpha_0^7$}; & & \node[block] (F) {$F$}; & \node[block] (Rep) {Rep}; & \node[rectangle] (brep) {}; \\
& & \node[block] (G0) {$G{\big|}\beta=0$}; & \node[block] (sign0) {Sign}; & \node[rectangle] (mux) {}; & \node[rectangle] (combine) {}; & \node[rectangle] (estimate) {$\beta_0^7$};\\
& & \node[block] (G1) {$G{\big|}\beta=1$}; & \node[block] (sign1) {Sign}; & \\
};

\node[rectangle] (breptext) at ($(brep.north)+(0,0.1cm)$) {\small $\beta$};

\node[connector] (c0) at (brep) {};
\node[connector] (c1) at ($(alpha)+(0.75cm,0)$) {};
\node[connector] (c2) at ($(G0.west)-(0.56cm,0)$) {};

\node[shape=mux2] (m0) at ($(mux)-(0,0.5cm)$) {};
\node[block, minimum height=2.5cm, minimum width = 0cm] (comb0) at (combine) {\rotatebox{90}{$Mix$}};

\draw[-] (alpha) -- (F);
\draw[-] (F) -- (Rep);
\draw[-] (Rep) -| ($(m0)+(0,0.675cm)$);

\draw[-] ($(alpha)+(0.75cm,0)$) |- (G0);
\draw[-] (G0) -- (sign0);
\draw[-] (sign0) -- (m0.in0|-sign0);

\draw[-] ($(alpha)+(0.75cm,0)$) |- (G1);
\draw[-] (G1) -- (sign1);
\draw[-] (sign1) -- (m0.in1|-sign1);

\draw[-] (Rep.east) -- (comb0.west|-Rep);
\draw[-] (m0.out) -- (comb0.west|-m0.out);
\draw[-] (comb0) -- (estimate);

\end{tikzpicture}
  }
  \caption{Architecture of the Rep1 Node.}
\label{fig:rep1_arch}
\end{figure}

Fig.~\ref{fig:rep1_arch} shows the architecture of the Rep1 node. It can be seen that there are two $G$ blocks. One preemptively calculates soft values assuming that the Rep block will output $\beta=0$ and the other for $\beta=1$. The Rep block provides a single bit estimate corresponding to the information bit the repetition code of length $N_v=4$ it is decoding. The outputs of the $G$ blocks go through a Sign block to generate hard decisions. The correct hard decision vector is then selected using the output of the Rep block. Finally, the bit estimate vector $\beta_0^7$ is built. The highest part, $\beta_4^7$, is always comprised of the multiplexer output. The lowest part, $\beta_0^3$, is either a copy of same output or its binary negation. The negated version is selected when the output of the Rep block is 1.

Calculations are carried out in one clock cycle. The output of the $F$, $G$ and Rep blocks are not stored in memory. Only the final result, the bit-estimate vector $\beta_0^7$, is stored in memory.

\subsection{High-Level Architecture}

\begin{figure}[t]
  \centering
  \resizebox{\columnwidth}{!}{
    \begin{tikzpicture}
\definecolor{deepgreen}{RGB}{8, 130, 25}

\usetikzlibrary{shapes,positioning,arrows,decorations.markings,fit}

\tikzset{
  block/.style={thick,draw,rectangle, minimum width = 3cm},
  >={latex},
  llr/.style={draw=red!50,fill=red!25},
  bit/.style={draw=blue!50,fill=blue!25},
  inst/.style={draw=deepgreen!50,fill=deepgreen!25},
}

\matrix[row sep=5mm, column sep=5mm] {

\node[block,llr] (alpha_ram) {$\alpha$-RAM}; & \node[block,llr] (chan_ram) {Channel RAM}; & \\
\node[block,llr] (alpha_router) {$\alpha$-Router}; & \node[block,llr] (chan_loader) {Channel Loader}; &  \node[rectangle] (chan) {Channel};\\
\node[block,very thick] (processor) {Processing Unit}; & \node[block] (controller) {Controller}; & \\
\node[block,bit] (beta_router) {$\beta$-Router}; & \node[block,inst] (inst_ram) {Instruction RAM};& \node[rectangle] (instructions) {Instructions}; \\
\node[block,bit] (beta_ram) {$\beta$-RAM}; & \node[block,bit] (codeword_ram) {Codeword RAM}; & \node[rectangle] (estimate) {Estimate};\\
};

\draw[thick,->] (chan) -- (chan_loader);

\draw[thick,->] (instructions) -- (inst_ram);
\draw[thick,->] (inst_ram) -- (controller);
\draw[thick,->] (controller) -- (processor);
\draw[thick,->] (controller) -- (chan_loader);
\draw[thick,->] (chan_loader) -- (chan_ram);

\draw[thick,<->] (alpha_router) -- (alpha_ram);
\draw[thick,->] (chan_ram) -- (alpha_router);
\draw[thick,<->] (alpha_router) -- (processor);
\draw[thick,<->] (beta_router) -- (processor);
\draw[thick,<->] (beta_router) -- (beta_ram);
\draw[thick,->] (beta_router) -- (codeword_ram);

\draw[thick,->] (codeword_ram) -- (estimate);

\end{tikzpicture}
  }
  \caption{High-level architecture of the decoder.}
\label{fig:impl:arch}
\end{figure}
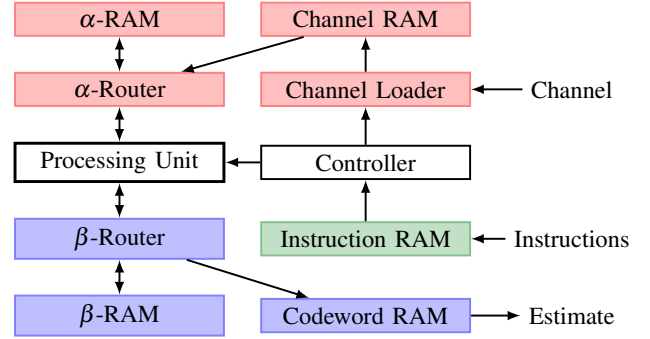

The high-level architecture of the decoder is presented in Fig.~\ref{fig:impl:arch}. Instructions representing the polar decoding operations to be performed are loaded before decoding starts. When the decoder is started, the controller signals the channel loader to start storing channel LLRs, 32 LLRs (160 bits) per clock cycle, into the channel RAM. The controller then starts to execute functions on the processing unit. The processing unit reads LLRs from the Channel or $\alpha$-RAM and writes LLRs to the $\alpha$-RAM. It reads or writes hard decisions to the $\beta$-RAM. The last $Combine$ operation writes the estimated codeword into the Codeword RAM, a memory accessible from outside the decoder.

The decoder is complete with all input and output buffers to accommodate loading a new frame and reading an estimated codeword while a frame is being decoded.
The required memory could be made smaller if the nominal throughput required is lower.
The loading or outputting of a full frame takes fewer clock cycles than the actual decoding, we have a pipelined operation; under normal operation, the decoder should not be slowed down by the I/O operations.

\subsection{Processing Unit or Processor}\label{sect:processor}

The core of the decoder is the processing unit illustrated in Fig.~\ref{fig:impl:processor} and based on the Fast-SSC implementation of \cite{Sarkis2014}. Thus, the processing unit features all the modules required to implement the nodes and operations described in sections \ref{sec:bg:fast-ssc} and \ref{sec:algo}. Notably, the 01 and RepSPC blocks connected to the $G$ block implement the 001 and 0RepSPC nodes, respectively, where the all-zero vector input is selected at the multiplexer $m_0$. The critical path of the decoder corresponds to the 0RepSPC node i.e. goes through $G$, RepSPC, the multiplexer $m_3$, $Combine$ and the multiplexer $m_2$. It is slightly longer than that of \cite{Sarkis2014}.

\begin{figure}[t]
  \centering
  \resizebox{\columnwidth}{!}{
    \definecolor{gcolor}{RGB}{155, 187, 89}
\definecolor{fcolor}{RGB}{79, 129, 189}
\definecolor{deepgreen}{RGB}{8, 130, 25}

\begin{tikzpicture}[every text node part/.style={align=center},
  every node/.style={font=\small}]

  \tikzset{every mux2 node/.style={draw,minimum width=0.5cm,minimum height=2cm,inner sep=1mm,outer sep=0pt}}
  \tikzset{every mux3 node/.style={draw,minimum width=0.5cm,minimum height=3cm,inner sep=1mm,outer sep=0pt}}
  \tikzset{every mux4 node/.style={draw,minimum width=0.5cm,minimum height=4cm,inner sep=1mm,outer sep=0pt}}
  \tikzset{every mux5 node/.style={draw,minimum width=0.5cm,minimum height=4.8cm,inner sep=1mm,outer sep=0pt}}

  \tikzset{
    block/.style={draw,rectangle, minimum width = 0.9cm, minimum height = 0.9cm},
    >={latex},
    connector/.style={circle,draw,fill=black,minimum size=3pt,inner sep=0},
  }

  \tikzstyle{critical}=[red,ultra thick]
  \tikzstyle{scritical}=[blue,very thick]

  \node[rectangle,inner sep=0pt, minimum width = 3mm] (alpha_in) at (0, 0) {$\mvec{\alpha}$};
  \node[rectangle,inner sep=0pt, minimum width = 3mm] (beta0_in) at (0, 1) {$\mvec{\beta_0}$};
  \node[rectangle,inner sep=0pt, minimum width = 3mm] (zero_in) at (0, 2) {$\mvec{0}$};

  \node[shape=mux2] (m0) at (0.85, 1.5) {$m_0$};

  \draw[-] (zero_in) --  (m0.in0);
  \draw[-] (beta0_in) -- (m0.in1);

  \node[block, minimum height = 2.5cm,very thick,gcolor] (g) at (2, 0.75) {$G$};
  \draw let \p1 = (g.west) in coordinate (g_in0) at (\x1, 1.5);
  \draw let \p1 = (g.west) in coordinate (g_in1) at (\x1, 0);

  \draw[-] (m0.out) -- coordinate (c6a) (g_in0);
  \draw[-] (alpha_in) -- coordinate (c9a) (g_in1) {};

  \node[rectangle,minimum width=0.5] (beta1_in) at (6.25, -2.175) {$\mvec{\beta_1}$};
  \node[block,very thick,orange] (spc) at (5.5, 0.325) {SPC};
  \node[coordinate] at ($(g.east) - (0, 0.425)$) (c3) {};
  \draw[-] (c3) -- coordinate (c4) ($(c3)+(0.3,0)$) -- (spc.west);
  \node[connector] at (c4) {};

  \node[block,very thick, purple] (0ml) at (4.4, -0.485) {Rep\\SPC};
  \draw[-] (c4) |- coordinate (c106) (0ml.west);
  \node[connector] at (c106) {};

  \node[block,very thick, blue,execute at begin node=\setlength{\baselineskip}{1em}] (0repspc) at (3.3, -1.35) {01};
  \draw[-] (c4) |- coordinate (c107) (0repspc.west);

  \node[mux5] (m2) at (9.8, 3.85) {$m_2$};
  \draw let \p1 = (m2.in0) in node[block, very thick,deepgreen] (rep) at (6.65, \y1) {Rep};
  \draw[-] (rep.east) |- (m2.in0);
  \draw let \p1 = (m2.in1) in node[block, very thick,deepgreen] (rep1) at (5.5, \y1) {Rep1};  
  \draw[-] (rep1.east) |- (m2.in1);
  \draw let \p1 = (m2.in2) in node[block,very thick, purple,execute at begin node=\setlength{\baselineskip}{1em}] (repspc) at (4.375, \y1) {Rep\\SPC};
  \draw[-] (repspc) |- (m2.in2);
  \node[block,very thick, blue] (ml) at (3.25, 3) {01};
  \draw[-] (ml.east) |- (m2.in3);

  \node[shape=mux5] (m3) at (7.5, -0.5) {$m_3$};

  \draw let \p1 = (m3.in0) in node[block,very thick] (sign) at (6.6, \y1) {Sign};
  \draw[-] (c4) |- coordinate (c5) (sign.west);
  \node[connector] at (c5) {};

  \draw[-] (sign.east) |- (m3.in0);
  \draw[-] (spc.east) |- (m3.in1);
  \draw[-] (0ml.east) |- (m3.in2);
  \draw[-] (0repspc.east) |- (m3.in3);
  \node[coordinate, left=0.375cm of m3.in4] (c1) {};
  \draw[-] (beta1_in) -| (c1);
  \draw[-] (c1) |- (m3.in4);

  \node[block, minimum height=4cm] (combine) at (8.5, 0.8) {\rotatebox{90}{$Combine$}};

  \node[coordinate] at ($(combine.west) - (0, 1.3)$) (c2) {};
  \draw[-] (m3.out) -- (c2);

  \node[coordinate, right=0.1cm of c6a] (c6) {};
  \node[coordinate, above=0.625cm of c6] (c7)  {};
  \draw[-] (c6) -- (c7);
  \node[connector] at (c6) {};
  \draw[-] let \p1 = (combine.west), \p2 = (c7) in (c7) -- (\x1, \y2) {};

  \node[block,very thick, fcolor] (f) at (5.5, 7.25) {$F$};
  \node[coordinate,right= 0.4cm of c9a] (c9) {};
  \node[connector] at (c9) {};
  \draw[-] (c9) |- (f.west);

  \draw let \p1 = (ml.west), \p2 = (c9) in coordinate (c10) at (\x2, \y1);
  \node[connector] at (c10) {};
  \draw[-] (c10) -- (ml.west);

  \draw let \p1 = (repspc.west), \p2 = (c9) in coordinate (c11) at (\x2, \y1);
  \node[connector] at (c11) {};
  \draw[-] (c11) -- (repspc.west);

  \draw let \p1 = (rep.west), \p2 = (c9) in coordinate (c12) at (\x2, \y1);
  \node[connector] at (c12) {};
  \draw[-] (c12) -- (rep.west);

  \draw let \p1 = (rep1.west), \p2 = (c9) in coordinate (c120) at (\x2, \y1);
  \node[connector] at (c120) {};
  \draw[-] (c120) -- (rep1.west);

  \node[mux2] (m1) at (7.75, 6.75) {$m_1$};
  \draw[-] (f.east) -- (m1.in0);
  \draw[-] (c5) |- (m1.in1);
  \node[circle, right=2.5cm of m1.east] (alpha_o) {$\mvec{\alpha'}$};
  \draw[-] (m1.out) -- (alpha_o);

  \node[coordinate, right=0.3cm of combine.east] (c13) {};
  \node[connector] at (c13) {};
  \draw[-] (combine.east) -- (c13);
  \draw[-] (c13) |- (m2.in4);

  \draw let \p1 = (alpha_o), \p2 = (m2.out) in node[circle] (beta0_o) at (\x1, \y2) {$\mvec{\beta_0'}$};
  \draw[-] (m2.out) -- (beta0_o);

  \draw let \p1 = (alpha_o), \p2 = (combine.east) in node[circle] (beta1_o) at (\x1, \y2) {$\mvec{\beta_1'}$};
  \draw[-] (combine.east) -- (beta1_o);

  \draw let \p1 = (combine.east), \p2 = (m2.in3) in node[coordinate] (c9) at (\x1, \y2) {};

\end{tikzpicture}
  }
  \caption{Architecture of the processing unit.}
\label{fig:impl:processor}
\end{figure}

\section{Results}
\label{sec:results}

\subsection{Verification Methodology}
A software model was used to generate random codewords for transmission using binary phase shift keying (BPSK) over an additive white Gaussian noise (AWGN) channel. The functionality of the designs was verified both at the RTL level and at the post-place and route level through simulations. Finally, the same frames were also decoded on an FPGA using an FPGA-in-the-loop setup. For all $\nicefrac{E_b}{N_0}$ values, a minimum of 100 frames in errors were simulated.

\subsection{Comparison with State-of-the-art Decoders}
\label{sec:vs-old-fast-ssc}

In this section, post-fitting results are presented for the Altera Stratix IV EP4SGX530KH40C2 FPGA. All results are worst-case using the slow 900 mV $85^\circ$C timing model. Table~\ref{tab:impl:results} shows the results for two rate-flexible implementations for polar codes of length 1024 and 2048, respectively. The decoder of \cite{Pamuk2013} is also included for comparison.

\begin{table}[t]
  \centering
  \caption{Post-fitting results for rate-flexible decoders for moderate-length polar codes.}
  \begin{tabular}{lcrrrc}
    \toprule
    \multirow{2}{*}{Implementation} & \multirow{2}{*}{$N$} & \multicolumn{1}{c}{\multirow{2}{*}{LUTs}} & \multicolumn{1}{c}{\multirow{2}{*}{Regs.}} & \multicolumn{1}{c}{RAM} & $f$ \\
    & & & & \multicolumn{1}{c}{\footnotesize(kbits)} & {\footnotesize(MHz)} \\
    \midrule
    \cite{Pamuk2013}    & 1024 &  1,940 &    748 &   7.1 & 239\vspace{2pt}\\
    \cite{Sarkis2014}*  & 1024 & 23,020 &  1,024 &  42.8 & 103 \\
                        & 2048 & 23,319 &  5,923 &  60.9 & 103\vspace{2pt}\\
    this work           & 1024 & 23,353 &  5,814 &  43.8 & 103 \\
                        & 2048 & 23,331 &  5,923 &  61.2 & 103 \\
    \bottomrule
  \end{tabular}
  \label{tab:impl:results}
\end{table}

Looking at the results for our proposed decoders, it can be observed that the number of look-up tables (LUTs) and registers required are very similar for both code lengths. However, the RAM usage differs significantly where decoding a longer code requires more memory as expected. Timing reports show that the critical path corresponds to the 0RepSPC node.

Table~\ref{tab:impl:results} also compares the proposed decoders against the decoder of \cite{Pamuk2013} as well as the original Fast-SSC implementation~\cite{Sarkis2014}. The latter was resynthesized so that the decoder only has to accommodate polar codes of length $N=1024$ or $N=2048$ and is marked with an asterisk (*) in Table~\ref{tab:impl:results}.

Our work requires at most a 1.4\% increase in used LUTs compared to~\cite{Sarkis2014}. The difference in registers can be mostly attributed to register duplication, a measure taken by the fitter to shorten the critical path to meet the requested clock frequency. The SRAM usage was also increased by 2.3\%.

\begin{table}[t]
  \centering
  \caption{Latency and information throughput comparison for low-rate moderate-length polar codes.}
  \begin{tabular}{lcccc}
    \toprule
    \multirow{3}{*}{Implementation} & \multirow{3}{*}{\shortstack{Code\\$(N,k)$}} & \multicolumn{2}{c}{Latency} & \multirow{3}{*}{\shortstack{Info. T/P\\(Mbps)}} \\
    \cmidrule(lr){3-4}
    &&(CCs)&($\mu$s)&\\
    \midrule
    \cite{Pamuk2013}     & $(1024,342)$ & 2185 & 9.14 & 37\\
                         & $(1024,512)$ & 2185 & 9.14 & 56\\
    \midrule
    \cite{Sarkis2014}*   & $(1024,342)$ & 201 & 1.95 & 175\\
                         & $(1024,512)$ & 220 & 2.14 & 240\\
                         & $(2048,683)$ & 366 & 3.55 & 192\\
                         & $(2048,1024)$& 389 & 3.78 & 271\vspace{2pt}\\
   \textit{altered codes}& $(1024,342)$ & 173 & 1.68 & 204\\
                         & $(1024,512)$ & 186 & 1.81 & 284\\
                         & $(2048,683)$ & 289 & 2.81 & 243\\
                         & $(2048,1024)$& 336 & 3.26 & 314\\
    \midrule
    this work            & $(1024,342)$ & 193 & 1.87 & 183\\
                         & $(1024,512)$ & 204 & 1.98 & 259\\
                         & $(2048,683)$ & 334 & 3.24 & 211\\
                         & $(2048,1024)$& 367 & 3.56 & 287\vspace{2pt}\\
   \textit{altered codes}& $(1024,342)$ & 157 & 1.52 & 224\\
                         & $(1024,512)$ & 165 & 1.60 & 320\\
                         & $(2048,683)$ & 274 & 2.66 & 257\\
                         & $(2048,1024)$& 308 & 2.99 & 342\\
    \bottomrule
  \end{tabular}
  \label{tab:impl:lat_and_tp}
\end{table}

Table~\ref{tab:impl:lat_and_tp} shows the latency and information throughput of the decoders of Table~\ref{tab:impl:results} when decoding low-rate moderate-length polar codes. It also shows the effect of using a polar codes with altered constructions---as described in Section~\ref{sec:code_construction}---with all Fast-SSC-based decoders. For both \cite{Sarkis2014}* and our work, the results listed as `altered codes' have the same resource usage and clock frequency as listed in Table~\ref{tab:impl:results} since these decoders can decode any polar code of length $N=1024$ or $N=2048$ by changing the code description in memory.

Applying the proposed altered construction alone, Table~\ref{tab:impl:lat_and_tp} shows that decoding these altered codes with the original decoders of \cite{Sarkis2014} results in a $14$\% to $21$\% latency reduction and a $16$\% to $27$\% throughput improvement. From the same table, it can be seen that decoding the unaltered codes with the updated hardware decoder integrating the proposed new constituent decoders, the latency is reduced by $4$\% to $10$\% and the throughput is improved by $4$\% to $10$\%. 

Combining the contribution of both the altered construction method and the new dedicated constituent decoders, the proposed work achieves the best latency among all compared decoders. For the polar codes of length $N=1024$, the throughput is $5.7$ to $6.1$ times greater than that of the two-phase decoder of \cite{Pamuk2013}. Finally, the latency is reduced by $22$\% to $28$\% and the throughput is increased by $26$\% to $34$\% over the Fast-SSC decoders of \cite{Sarkis2014}. 

Table~\ref{tab:cmp_asic} presents a comparison of this work against the state-of-the-art ASIC implementations. Our ASIC results are for the 65~nm CMOS GP technology from TSMC and are obtained with Cadence RTL Compiler. Only registers were used for memory due to the lack of access to an SRAM compiler. Normalized results for the decoders from the literature are also provided. For consistency, only results for a $(1024, 512)$ polar code are compared to match what was done in the other works. It should be noted that \cite{Park2014} provides measurement results.

\begin{table}
  \centering
  \setlength{\tabcolsep}{1.5pt}
  \caption{Comparison of state-of-the-art ASIC decoders decoding a (1024, 512) polar code.}
  \begin{tabular}{l c c c c c}
    \toprule
                 &\multicolumn{2}{c}{\bf This work}& \phantom{$^\diamond$}\cite{Park2014}$^\diamond$ & \cite{Dizdar2015} & \cite{Yuan2014} \\
    \midrule
    Algorithm    & \multicolumn{2}{c}{Fast-SSC} & BP              & SC    & 2-bit SC\\
    Technology   &    \multicolumn{2}{c}{65 nm} & 65 nm           & 90 nm & 45 nm\\
    Supply (V)   &           0.8 &          1.0 & 1.0             & 1.3   & N/A \\
    Oper. temp. ($^\circ$C)&   25 &           25 & $\approx 25$    & N/A   & N/A \\
    Area (mm$^2$)&          0.69 &         0.69 & 1.48            & 3.21  & N/A \\
    Area @65nm (mm$^2$)&    0.69 &         0.69 & 1.48            & 1.68  & 0.4\\
    Frequency (MHz)&         400 &          600 & 300             & 2.5   & 750\\
    Latency ($\mu$s)&       0.41 &         0.27 & 50              & 0.39  & 1.02\\
    Info. T/P (Gbps)&       1.24 &         1.86 & 2.4 @ 4dB       & 1.28  & 0.5\\
    Sust. Info. T/P (Gbps)& 1.24 &         1.86 & 1.0             & 1.28  & 0.5\\
    Area Eff. (Gbps/mm$^2$)& 1.8 &          2.7 & 1.6 @ 4dB       & 0.4   & N/A\\
    Power (mW)   &            96 &          215 & 478             & 191   & N/A\\ 
    Energy (pJ/bit)&          77 &          115 & 203 @ 4dB       & 149   & N/A \\ 
    \bottomrule
    &&&\vspace{-6pt}\\
    \multicolumn{6}{l}{\textit{$\diamond$ Measurement results.}}\\
  \end{tabular}
  \label{tab:cmp_asic}
\end{table}

From Table~\ref{tab:cmp_asic}, it can be seen that both implementations of our proposed decoder---at different supply voltages---are 46\% and 42\% the size of the BP decoder \cite{Park2014} and the combinational decoder \cite{Dizdar2015}, respectively, when all areas are normalized to 65nm technology. Our work has two orders of magnitude lower latency than the BP decoder of \cite{Park2014}, and two to five times lower latency than \cite{Yuan2014}. The latency of the proposed design is 1.05 times and 0.7 times that of \cite{Dizdar2015}, when operating at 400 and 600 MHz, respectively. The BP decoder \cite{Park2014} employs early termination and its throughput at $\nicefrac{E_b}{N_0} = 4$ dB is the fastest followed by our proposed design. Since the area reported in \cite{Park2014} excludes the memory necessary to buffer additional received vectors to sustain the variable decoding latency due to early termination, we also report the sustained throughput for that decoder. The sustained throughput is 1.0 Gbps as a maximum of 15 iterations is required for the BP decoder to match the error-correction performance of the SC-based decoders.
Comparing the information throughput of all decoders---using the best-case values for BP,--- it can be seen that the area efficiency of our decoder is the greatest. Lastly, the power consumption estimations indicate that our decoders are more energy efficient than the BP decoder of \cite{Park2014}. Our proposed decoders are also more energy efficient than that of \cite{Dizdar2015}. However, due to the difference in implementation technology, the results of this latter comparison could change if \cite{Dizdar2015} were to be implemented in 65nm.

\section{Conclusion}
\label{sec:conclusion}
In this work, we showed how the original Fast-SSC algorithm implementation could be improved by adding dedicated decoders for three new types of constituent codes frequently appearing in low-rate codes. We also used polar code construction alterations to significantly reduce the latency and increase the throughput of a Fast-SSC decoder at the cost of a small error-correction performance loss. Rate-flexible polar decoders for polar codes of lengths 1024 and 2048 were implemented on an FPGA. Four low-rate polar codes with competitive error-correction performance were proposed. Their resulting latency and throughput represent a $22$\% to $28$\% reduction and a $26$\% to $34$\% improvement over the previous work, respectively. The information throughput was shown to be $224$, $320$, $257$, and $342$ Mbps at approximately $100$~MHz on the Altera Stratix IV FPGAs for the $(1024,342)$, $(1024,512)$, $(2048,683)$ and $(2048,1024)$ polar codes, respectively. On 65~nm ASIC CMOS technology, the proposed decoder for a $(1024, 512)$ polar code was shown to compare favorably against the state-of-the-art ASIC decoders. With a clock frequency of 400 MHz and a supply voltage of 0.8 V, it has a latency of 0.41 $\mu$s and an area efficiency of 1.8 Gbps/mm$^2$ for an energy efficiency of 77 pJ/info. bit. At 600 MHz with a supply of 1 V, the latency is reduced to 0.27 $\mu$s and the area efficiency increased to 2.7 Gbps/mm$^2$ at 115 pJ/info. bit.

\section*{Acknowledgement}
This work was supported by the Natural Sciences and Engineering Research Council of Canada (NSERC). Claude Thibeault is a member of ReSMiQ. Warren J. Gross is a member of ReSMiQ and SYTACom.

\bibliographystyle{IEEEtran}
\bibliography{IEEEabrv,refs}

\end{document}